# Comprehensive Wide-Band Magnitudes and Albedos for the Planets, With Applications to Exo-Planets and Planet Nine

Anthony Mallama

14012 Lancaster Lane, Bowie, MD, 20715, USA

anthony.mallama@gmail.com

Bruce Krobusek

5950 King Hill Drive, Farmington, NY, 14425, USA

bkrobusek@gmail.com

Hristo Pavlov

9 Chad Place, St. Clair, NSW 2759, Australia

hristo_dpavlov@yahoo.com

Abstract

Complete sets of reference magnitudes in all 7 Johnson-Cousins bands (U, B, V, R, I, $R_C$ and $I_C$) and the 5 principal Sloan bands (u', g', r', i', and z') are presented for the 8 planets. These data are accompanied by illumination phase functions and other formulas which characterize the instantaneous brightness of the planets. The main source of Johnson-Cousins magnitudes is a series of individualized photometric studies reported in recent years. Gaps in that dataset were filled with magnitudes synthesized in this study from published spectrophotometry. The planetary Sloan magnitudes, which are established here for the first time, are an average of newly recorded Sloan filter photometry, synthetic magnitudes and values transformed from the Johnson-Cousins system. Geometric albedos derived from these two sets of magnitudes are consistent within each photometric system and between the systems for all planets and in all bands. This consistency validates the albedos themselves as well as the magnitudes from which they were derived. In addition, a quantity termed the *delta stellar magnitude* is introduced to indicate the difference between the magnitude of a planet and that of its parent star. A table of these delta values for exo-planets possessing a range of physical characteristics is presented. The delta magnitudes are for phase angle $90^o$ where a planet is near the greatest apparent separation from its star. This quantity may be useful in exo-planet detection and observation strategies when an estimate of the signal-to-noise ratio is needed. Likewise, the phase curves presented in this paper can be used for characterizing exo-planets. Finally, magnitudes for the proposed Planet Nine are estimated, and we note that P9 may be especially faint at red and near-IR wavelengths.

1. Introduction

The *absolute* magnitude for a planet is that corresponding to full solar illumination of the disk when the observer and the Sun are at a distance of one AU. Many astronomical reference volumes, textbooks and journal articles list characteristic values of the absolute planetary magnitudes which are referred to herein as *reference* magnitudes. These may represent an average of absolute magnitudes over time or a mean value derived from different sources. The absolute and reference magnitudes are distinct quantities because the planets are intrinsically variable in brightness.

Observed magnitudes depend strongly on the *illumination phase angle* which is that subtended at the planet by the Sun and the observer. The magnitude at any phase may be evaluated by adjusting the reference magnitude according to an illumination *phase function* derived from observations spanning a range of phase angles. Phases can also refer to other aspects of a planet besides illumination, such as the rotation angle around the polar axis, which may influence the observed magnitude. Geophysical information can be derived from the analysis of planetary phase functions and by characterizing the varying absolute magnitudes over time. Some aspects of exo-planets can also be inferred by similar methods.

Many of the magnitudes listed in reference volumes, almanacs and textbooks do not take account of results published in the recent literature or they are lacking data for certain planets and for certain band-passes. Furthermore, none of these works contain the Sloan magnitudes because those are being reported here for the first time. The purpose of this paper, therefore, is to provide a complete list of reference magnitudes for the two principal wide-band photometric regimes in use today, namely, the Johnson-Cousins system (Johnson et al. 1966, Cousins, 1976a and Cousins 1976b) and the Sloan system (Smith et al., 2002 and Fukugita et al., 1996).

Section 2 characterizes the two photometric systems. Section 3 briefly reviews the published literature on Johnson-Cousins photometric planetary magnitudes, describes how synthetic magnitudes were derived from spectrophotometry and lists the adopted reference magnitudes. Section 4 describes the new Sloan photometry which was performed for this study. Section 5 lists the Sloan photometric results along with magnitudes derived synthetically and by transforming from Johnson-Cousins values. Sloan reference magnitudes are then reported. Section 6 lists geometric albedos and demonstrates the consistency between the Johnson-Cousins and Sloan results. The good agreement between the albedos on the two systems validates the data for both data sets and the magnitudes as well. Section 7 discusses how these results may be applied to exo-planet research. Estimated magnitudes for the proposed Planet Nine are given in Section 8 and, finally, Section 9 summarizes the paper.

Much of the work reported here involved detailed analyses of photometric observations as well as interpretation of extensive sets of brightness phase functions. Such complexities would only interest a few specialists, so those details have been placed in an appendix. A reader looking for reference magnitudes along with a general explanation of how they were derived will find that material in the body of the paper.

2. The photometric systems

The Johnson system was originally defined by Johnson et al. (1966) while the Cousins extension was described ten years later (Cousins, 1976a and 1976b). The combined system spans wavelengths from about 0.3 to 1.0 μm and contains the 7 band-passes, U, B, V, R, I, $R_C$ and $I_C$, characterized in Table 1.

Table 1: Characteristics of the Johnson-Cousins photometric system

| Filter | Effective Wavelength (μm) | Full-width half-max. (μm) |
|--------|---------------------------|---------------------------|
| U      | 0.360                     | 0.068                     |
| B      | 0.436                     | 0.098                     |
| V      | 0.549                     | 0.086                     |
| R      | 0.700                     | 0.209                     |
| I      | 0.900                     | 0.221                     |
| Rc     | 0.641                     | 0.158                     |
| Ic     | 0.798                     | 0.154                     |

The Sloan photometric system has gained widespread acceptance in the astronomical community during recent years. There are five primary bands, u, g, r, i, and z, in the Sloan Digital Sky Survey which cover approximately the same total wavelength range as does the Johnson-Cousins system. However the individual band-passes differ significantly as shown in Table 2. Magnitudes designated with primes, u', g', r', i', and z', indicate the system established by Smith et al. (2002) that includes 158 standard stars with Sloan magnitudes and which is preferred for photometry. While stellar magnitudes are being reported on this new system with increasing frequency no equivalent values for the planets have been published until now.

The Sloan filters represent a marked improvement over their Johnson-Cousins counterparts. While the latter are merely colored glass, the Sloan filters have an added dielectric coating to steepen the shoulders of the response curves. Thus, the Sloan filters exhibit less overlap between adjacent bands and have more rectangular response characteristics than the Johnson-Cousins filters. Furthermore, the emission line from the terrestrial sky at 0.558 μm lies between the g' and r' filters on the Sloan system. A final advantage of the Sloan system is that its magnitudes are directly related to absolute flux, that is, they are on the AB system of Oke and Gunn (1983). In addition to the SDSS itself, the Large Synoptic Survey Telescope (LSST) and the Panchromatic Survey Telescope and Rapid Response System (Pan-Starrs) also use Sloan filters.

Table 2: Characteristics of the Sloan photometric system

| Filter | Effective Wavelength (μm) | Full-width half-max. (μm) |
|---|---|---|
| u' | 0.355 | 0.063 |
| g' | 0.469 | 0.143 |
| r' | 0.616 | 0.140 |
| i' | 0.748 | 0.149 |
| z' | 0.893 | 0.117 |

## 3. Johnson-Cousins magnitudes

Several compilations of UBV planetary magnitudes were published around 50 years ago. Those by Harris (1961), de Vaucouleurs (1964) and Irvine et al. (1968) are especially notable. Furthermore, Lockwood, Jerzykiewicz and their colleagues have monitored the narrow band magnitudes of several solar system bodies from the mid-twentieth century until the present (e.g., Lockwood and Jerzykiewicz, 2006).

The wide-band results have all been updated by individual brightness studies for every planet except the Earth during the past 14 years. These recent studies benefit from extensive monitoring of a great variety of phase functions which have been found to affect planetary magnitudes. They also include results from more instruments measuring more band-passes over a longer period of time. The papers published since 2002 are briefly recounted in the next paragraph while more details are available in the appendix.

Mercury (Mallama et al. 2002) was observed with the Solar and Heliospheric Observatory satellite and also with the ground-based instrumentation. The planet exhibits a very strong brightness surge when fully illuminated due to coherent backscattering from its regolith. Analysis of Mercury's illumination phase curve provided a measure of its surface roughness which was found to be similar to that of the moon. Venus (Mallama et al. 2006) was also observed with SOHO and from the ground. An anomalous peak in the phase curve near phase $170^o$ was modeled by forward scattering of sunlight from droplets of sulfuric acid in the planet's upper atmosphere. Studies of the planets beyond the Earth's orbit utilized ground-based observations spanning tens of years and covering a wide range of planetary aspects. The brightness of Mars (Mallama 2007) depends on its orbital and rotational phase angles in addition to the illumination phase. Global dust storms were also found to produce a 15% brightness excess. The magnitude of Jupiter (Mallama and Schmude 2012) is fairly constant although changing intensities of its cloud belts produce measurable variations of the integrated flux. The inclination of Saturn's rings (Schmude 2011 and Mallama 2012) contributes to a brightness variation approaching two magnitudes. Uranus (Schmude et al. 2015) also varies as a function of its inclination, especially at red and near-IR wavelengths. The brightness of this planet increases when its methane-depleted polar regions are tilted toward the Earth and Sun. Neptune (Schmude et al 2016 ) has brightened substantially since the 1980s though the mechanism for this change is still not understood.

Photometric reference magnitudes from the studies summarized above are listed in Table 3. The photometric data and derived results from the papers cited above have been used extensively by researchers studying the geophysical and astronomical aspects of the planets. Some of the references to our papers are listed in the appendix under the sections on individual planets.

The magnitude of a planet can be synthesized by integrating the product of its flux and the instrumental response of a standard photometric band over the frequency range of that band. The relationship is indicated by Equation 1 (taken from Smith et al., 2002 and Fukugita et al., 1996), where $v$ is frequency, $S_v$ is the system response and $f_v$ is flux in ergs/s/cm$^2$/Hz at the top of the atmosphere. The values of $S_v$ were taken from the Johns Hopkins Filter Profile Service at http://skyservice.pha.jhu.edu and from http://spiff.rit.edu/classes/phys440/lectures/filters/filters.html.

$$m = -2.5 \frac{\int d(\log v) \, f_v \, S_v}{\int d(\log v) \, S_v}$$

Equation 1

Spectrophotometric data for each planet was located in the published literature. In some cases flux values in energy units had to be derived from albedos and similar non-energy measurements. Whole disk results were available for most of the planets, but they were deduced from disk-resolved data for Mercury and Venus. The sources of spectrophotometric data and more detailed descriptions of the synthesis for each individual planet are given in the appendix.

The photometric and synthetic Johnson-Cousins magnitudes for planets are reported in Table 3. When both magnitudes are listed, the photometric value is adopted as the reference value and its formal uncertainty is also given where available. When no photometric magnitude is available the synthetic magnitude is adopted. Preference is given to photometric results because they correspond to data collected by multiple instruments over years or decades of observation. The synthetic magnitudes, on the other hand, are derived from fewer sources of data spanning shorter periods of time.

Table 3. Johnson-Cousins magnitudes

| Planet  | Method      | U     | B     | V     | R     | I     | Rc    | Ic    |
|---------|-------------|-------|-------|-------|-------|-------|-------|-------|
| Mercury | Photometric | ----  | ----  | -0.69 | ----  | ----  | ----  | ----  |
| "       | Synthetic   | 0.69  | 0.28  | -0.68 | -1.44 | -1.99 | -1.21 | -1.69 |
| "       | Reference   | 0.69  | 0.28  | -0.69 | -1.44 | -1.99 | -1.21 | -1.69 |
| "       | Uncertainty | ----  | ----  | 0.03  | ----  | ----  | ----  | ----  |
| Venus   | Photometric | ----  | -3.68 | -4.38 | -4.95 | -5.08 | ----  | ----  |
| "       | Synthetic   | -2.79 | -3.46 | -4.38 | -4.91 | ----- | -4.73 | -5.04 |
| "       | Reference   | -2.79 | -3.68 | -4.38 | -4.95 | -5.08 | -4.73 | -5.04 |
| "       | Uncertainty | ----  | ----  | ----  | ----  | ----  | ----  | ----  |
| Earth   | Photometric | ----  | ----  | ----  | ----  | ----  | ----  | ----  |
| "       | Synthetic   | -3.64 | -3.52 | -3.99 | -4.49 | -4.86 | -4.28 | -4.63 |
| "       | Reference   | -3.64 | -3.52 | -3.99 | -4.49 | -4.86 | -4.28 | -4.63 |
| "       | Uncertainty | ----  | ----  | ----  | ----  | ----  | ----  | ----  |
| Mars    | Photometric | 0.39  | -0.24 | -1.60 | -2.71 | -3.20 | ----  | ----  |
| "       | Synthetic   | 0.59  | -0.04 | -1.54 | -2.67 | -3.16 | -2.42 | -2.90 |
| "       | Reference   | 0.39  | -0.24 | -1.60 | -2.71 | -3.20 | -2.42 | -2.90 |
| "       | Uncertainty | 0.01  | 0.01  | 0.01  | 0.01  | 0.01  | ----  | ----  |
| Jupiter | Photometric | -8.11 | -8.54 | -9.40 | -9.85 | -9.72 | ----  | ----  |
|         | Synthetic   | -8.14 | -8.54 | -9.42 | -9.89 | -9.85 | -9.75 | -9.79 |
| "       | Reference   | -8.11 | -8.54 | -9.40 | -9.85 | -9.72 | -9.75 | -9.79 |
| "       | Uncertainty | 0.01  | 0.01  | 0.01  | 0.01  | 0.01  | ----  | ----  |
| Saturn  | Photometric | -7.08 | -7.84 | -8.91 | -9.59 | -9.61 | ----  | ----  |
| "       | Synthetic   | -7.32 | -8.02 | -9.08 | -9.76 | -9.83 | -9.59 | -9.74 |
| "       | Reference   | -7.08 | -7.84 | -8.91 | -9.59 | -9.61 | -9.59 | -9.74 |
| "       | Uncertainty | ----  | ----  | ----  | ----  | ----  | ----  | ----  |
| Uranus  | Photometric | -6.28 | -6.61 | -7.11 | -6.69 | ----  | -6.84 | -6.00 |
| "       | Synthetic   | -6.40 | -6.59 | -7.16 | -6.80 | -6.01 | -6.81 | -6.11 |
| "       | Reference   | -6.28 | -6.61 | -7.11 | -6.69 | -6.01 | -6.84 | -6.00 |
| "       | Uncertainty | 0.05  | 0.02  | 0.02  | 0.02  | ----  | 0.05  | 0.05  |
| Neptune | Photometric | -6.38 | -6.55 | -6.94 | -6.50 | ----  | -6.61 | -5.71 |
| "       | Synthetic   | -6.43 | -6.52 | -6.95 | -6.70 | -5.77 | -6.67 | -5.78 |
| "       | Reference   | -6.38 | -6.55 | -6.94 | -6.50 | -5.77 | -6.61 | -5.71 |
| "       | Uncertainty | 0.05  | 0.02  | 0.02  | 0.05  | ----  | 0.08  | 0.08  |

4. Sloan system photometry

This section describes new photometry recorded on the Sloan system. We discuss the standard stars used for brightness references, the optical and sensor hardware employed, the observing procedures and the methods of calibration.

The Sloan standard stars of Smith et al. (2002) range from about magnitude r' = 9 to r' = 14 and, thus, are generally too faint to serve as references for planetary photometry. So, in preparation for this study, Mallama (2014) generated a catalog of bright secondary Sloan standards in the range -1 < r' < 7 by transforming the magnitudes of standard stars of the Johnson-Cousins system. Validation was accomplished by performing photometry on a sample of the catalog stars and also by deriving synthetic magnitudes from Hubble Space Telescope spectra for a separate sample (Mallama and Krobusek, 2015). The uncertainty of the secondary standards is estimated to be 0.03 magnitude in the g', r', i' and z' filters, while it is about 0.08 in the ultraviolet u' band-pass. Chonis and Gaskell (2008) and others have noted the difficulties in transforming between Johnson-Cousins U and Sloan u'.

Author BK obtained the photometric measurements reported in this paper with a 20-cm aperture Schmidt-Cassegrain telescope, an SBIG CCD camera containing a cooled Kodak KAF-0400 sensor, and a set of five Generation 2 Astrodon Sloan filters. The observations were scheduled so that the air masses were generally less than 1.5. Reference stars were chosen to be nearby their planets or at least to have similar elevations at the time of observation so that the differential air masses were usually less than 0.1. These precautions were taken in order to minimize uncertainties due to atmospheric extinction.

As a further precaution, the approximate magnitude and color of each planet was taken into account when selecting its standard star. The matching of magnitudes allowed for a similarly strong signal-to-noise ratio for both the planet and its standard, while color matching minimized errors in the color transformation step of calibration.

The procedure for acquiring image data for a single Sloan magnitude in a single filter was to record 2 separate series of 3 images of the planet interleaved with 2 such series for the standard star. The resulting magnitude for that filter represents the average of 6 values derived from 6 pairs of planet-and-standard CCD images. Thus, each set of 5 u', g', r', i', z' magnitudes derives from 60 separate images. Additional flat field and dark frames were recorded at all observing sessions.

Planetary photometry differs from stellar photometry in at least three ways. The planets are extended objects, some of them are extremely bright, and the two innermost planets should be observed during daylight when they are high in the sky in order to minimize atmospheric extinction. The special observational procedures used to address these unique challenges are discussed for each planet in the appendix.

The instrumental magnitudes derived from reduced CCD images were corrected for extinction and transformed to standard Sloan magnitudes. The methods originally developed by Hardie (1962) for the UBV system were adapted to the Sloan system and his technique was enhanced to include error estimation. The final error estimate accounts for the standard deviation of the mean of the six magnitudes determined for each filter, in addition to error propagation due to the uncertainties in the extinction and transformation coefficients. The observational results are listed in the appendix.

The methods used to characterize the color response of the hardware and the effect of atmospheric extinction at the observing site were also adapted from those of Hardie. Color transformation coefficients were derived from observations of the primary standard stars taken at small air masses. Atmospheric extinction coefficients were determined from time-series observations of secondary standard stars as they traversed a range of about one air mass at the observing site in Farmington, NY, USA. The resulting coefficients are listed in the Table 4.

Table 4. Sloan photometric calibration coefficients

| Band | Extinction* | Transformation** |
|------|-------------|------------------|
| u'   | 0.69        | +0.102           |
| g'   | 0.31        | +0.064           |
| r'   | 0.17        | +0.019           |
| i'   | 0.10        | −0.072           |
| z'   | 0.08        | −0.017           |

```
 * Magnitudes per air mass. Second order extinction was taken
   to be -0.02 for u' and g', and zero otherwise.
** Color indices corresponding to the transformation
   coefficient are as follows: u'-g' (u'), g'-r' (g'), g'-r'
   (r'), r'-i' (i') and r'-z' (z').
```

The absolute magnitude for a planet was defined in the introductory section as that corresponding to full solar illumination of the disk with the observer and the Sun at a distance of one AU. Equation 2 indicates that value as *M₁(0)* where the '*1*' indicates one AU and '*0*' indicates that the phase angle, $\alpha$, is zero. *M₁($\alpha$)* is the magnitude when $\alpha$ is not zero and thus the brightness is diminished in accordance with the illumination phase function which is represented by the following polynomial (Harris, 1961).

$$M_1(\alpha) = M_1(0) + C_1 \alpha + C_2 \alpha^2 + \ldots$$

Equation 2

The values of $M_1(0)$ derived in this study depend on the observed values of $M_1(\alpha)$ and the polynomial coefficients, listed in the appendix. The polynomials were based upon phase angle coefficients measured in the Johnson-Cousins magnitude system and transformed to the Sloan system, also as indicated in the appendix. The functions include additional phase relations besides the solar illumination, such as the rotation angle and the season, for some planets. This prior knowledge of the planetary phase functions allowed for values of $M_1(0)$ to be determined from a set of Sloan observations recorded over just two years. Otherwise many decades would have been needed to sample all the possible aspects of planetary geometry and geophysics.

5. Sloan magnitudes

The averaged values of $M_1(0)$ for each planet are listed in Table 5 in addition to synthesized and transformed Sloan magnitudes. The synthesized magnitudes were derived in exactly the same way as those of the Johnson-Cousins values described in Section 3. However, the system response functions were those referenced by Smith et al. (2002) for the Sloan system and retrieved from http://www-star.fnal.gov/ugriz/Filters/response.html.

The UBVRI magnitudes of the planets listed in Table 3 were transformed to Sloan magnitudes by the same procedure used to generate magnitudes for the secondary reference stars (Mallama, 2014). An intermediate step depends on a star's spectral type and its luminosity class (Taylor, 1986). Since the spectral energy distributions of the planets do not correspond exactly with those of stars, spectral classes were approximated based on the planetary B-V color index. The luminosity class was taken to be the main sequence. The available sources of data were weighted equally in determining the reference magnitudes. The uncertainty is the standard deviation of the mean of the values used to compute the reference magnitudes.

Table 5. Sloan magnitudes

| Planet  | Method      | u'    | g'    | r'    | i'    | z'    |
|---------|-------------|-------|-------|-------|-------|-------|
| Mercury | Photometric | 1.37  | -0.44 | -1.11 | -1.41 | -1.56 |
| "       | Synthetic   | 1.62  | -0.21 | -0.96 | -1.24 | -1.39 |
| "       | Transformed | 1.43  | -0.25 | -1.01 | -1.30 | -1.46 |
| "       | Reference   | 1.47  | -0.30 | -1.03 | -1.32 | -1.47 |
| "       | Uncertainty | 0.09  | 0.09  | 0.06  | 0.06  | 0.08  |
| Venus   | Photometric | -1.88 | -4.07 | -4.57 | -4.66 | -4.54 |
| "       | Synthetic   | -1.89 | -3.98 | -4.52 | -4.62 | ----- |
| "       | Transformed | -1.75 | -4.09 | -4.58 | -4.71 | -4.47 |
| "       | Reference   | -1.84 | -4.04 | -4.56 | -4.66 | -4.51 |
| "       | Uncertainty | 0.08  | 0.04  | 0.02  | 0.03  | 0.04  |
| Earth   | Photometric | ----- | ----- | ----- | ----- | ----- |
| "       | Synthetic   | -2.77 | -3.85 | -4.06 | -4.22 | -4.23 |
| " | Transformed | -2.86 | -3.83 | -3.96 | -4.05 | -4.08 |
| " | Reference | -2.82 | -3.84 | -4.01 | -4.14 | -4.16 |
| " | Uncertainty | 0.06 | 0.01 | 0.07 | 0.12 | 0.11 |
| | | | | | | |
| Mars | Photometric | 1.21 | -0.85 | -2.18 | -2.54 | -2.57 |
| " | Synthetic | 1.50 | -0.69 | -2.13 | -2.49 | -2.53 |
| " | Transformed | 1.04 | -0.95 | -2.11 | -2.35 | -2.48 |
| " | Reference | 1.25 | -0.84 | -2.14 | -2.46 | -2.53 |
| " | Uncertainty | 0.16 | 0.09 | 0.03 | 0.07 | 0.03 |
| | | | | | | |
| Jupiter | Photometric | -7.30 | -9.09 | -9.65 | -9.59 | -9.23 |
| " | Synthetic | -7.25 | -9.02 | -9.60 | -9.50 | -9.08 |
| " | Transformed | -7.31 | -9.02 | -9.60 | -9.33 | -9.14 |
| " | Reference | -7.29 | -9.04 | -9.62 | -9.47 | -9.15 |
| " | Uncertainty | 0.02 | 0.03 | 0.02 | 0.09 | 0.05 |
| | | | | | | |
| Saturn | Photometric | -6.00 | -8.40 | -9.25 | -9.30 | -9.01 |
| " | Synthetic | -6.45 | -8.57 | -9.38 | -9.43 | -9.06 |
| " | Transformed | -6.26 | -8.42 | -9.23 | -9.10 | -8.98 |
| " | Reference | -6.24 | -8.46 | -9.28 | -9.27 | -9.02 |
| " | Uncertainty | 0.16 | 0.07 | 0.06 | 0.12 | 0.03 |
| | | | | | | |
| Uranus | Photometric | -5.57 | -6.97 | -6.82 | -5.76 | -4.91 |
| " | Synthetic | -5.54 | -6.97 | -6.92 | -5.92 | -5.05 |
| " | Transformed | -5.37 | -6.93 | -6.97 | -5.82 | ----- |
| " | Reference | -5.49 | -6.96 | -6.91 | -5.83 | -4.98 |
| " | Uncertainty | 0.08 | 0.02 | 0.05 | 0.06 | 0.11 |
| | | | | | | |
| Neptune | Photometric | -5.40 | -6.86 | -6.58 | -5.51 | -4.71 |
| " | Synthetic | -5.57 | -6.86 | -6.71 | -5.68 | -4.79 |
| " | Transformed | -5.46 | -6.82 | -6.77 | -5.59 | ----- |
| " | Reference | -5.48 | -6.85 | -6.69 | -5.60 | -4.74 |
| " | Uncertainty | 0.06 | 0.02 | 0.07 | 0.06 | 0.07 |

6. Geometric albedos

A planet's geometric albedo is defined as the ratio of its observed flux to that of a perfectly reflecting Lambertian disk. To give an example, the V magnitude of Saturn is -8.91 from Table 3 and the V magnitude of the Sun is −26.75 from Table 6. So, the ratio of the luminosity of Saturn to that of the Sun is $7.31 \times 10^{-8}$ as shown in Equation 3.

$$Lratio = 10^{\{(-26.75+8.91)/2.5\}} = 7.31 \times 10^{-8}$$

Equation 3

The average Saturnian disk radius, *r*, is 57,240 km including oblateness. The AU distance is $149.6 \times 10^6$ km, hence there is an area factor, $\sin^2(r/AU)$ or $1.46 \times 10^{-7}$. After accounting for area, the geometric albedo, *p*, is 0.499 as shown in Equation 4.

$$p = Lratio / \sin^2(r/AU) = 0.499$$

Equation 4

The comprehensive list of albedos from Table 7 is plotted in Fig. 1. The agreement between the Sloan and Johnson-Cousins values for all eight planets and over all wavelengths serves as validation for the albedos as well as the magnitudes from which they were derived.

Table 6. Solar magnitudes

```
        U*       B*      V*       R*       I*      R_C**    I_C**
      -25.90  -26.10  -26.75   -27.29   -27.63   -27.15   -27.49

               u'       g'       r'       i'       z'
             -25.02   -26.45  -26.89   -27.00   -26.97

 * Livingston (2001)
 ** Binney and Merrifield (1998)
 Sloan values retrieved from https://www.astro.umd.edu/~ssm/ASTR620/mags.html
```

Table 7. Geometric albedos

Johnson-Cousins:

| Planet | U | B | V | R | I | $R_C$ | $I_C$ |
|---|---|---|---|---|---|---|---|
| Mercury | 0.087 | 0.105 | 0.142 | 0.172 | 0.208 | 0.158 | 0.180 |
| Venus | 0.348 | 0.658 | 0.689 | 0.708 | 0.584 | 0.658 | 0.640 |
| Earth | 0.688 | 0.512 | 0.434 | 0.418 | 0.430 | 0.392 | 0.396 |
| Mars | 0.060 | 0.088 | 0.170 | 0.288 | 0.330 | 0.250 | 0.285 |
| Jupiter | 0.358 | 0.443 | 0.538 | 0.495 | 0.321 | 0.513 | 0.389 |
| Saturn | 0.203 | 0.339 | 0.499 | 0.568 | 0.423 | 0.646 | 0.543 |
| Uranus | 0.502 | 0.561 | 0.488 | 0.202 | 0.079 | 0.264 | 0.089 |
| Neptune | 0.578 | 0.562 | 0.442 | 0.181 | 0.067 | 0.226 | 0.072 |

Sloan:

| Planet | u' | g' | r' | i' | z' |
|---|---|---|---|---|---|
| Mercury | 0.095 | 0.130 | 0.169 | 0.200 | 0.237 |
| Venus | 0.326 | 0.664 | 0.712 | 0.708 | 0.630 |
| Earth | 0.722 | 0.497 | 0.388 | 0.393 | 0.412 |
| Mars | 0.061 | 0.111 | 0.245 | 0.298 | 0.325 |
| Jupiter | 0.377 | 0.509 | 0.575 | 0.456 | 0.348 |
| Saturn | 0.209 | 0.436 | 0.618 | 0.601 | 0.450 |
| Uranus | 0.541 | 0.559 | 0.355 | 0.120 | 0.056 |
| Neptune | 0.564 | 0.533 | 0.307 | 0.101 | 0.048 |

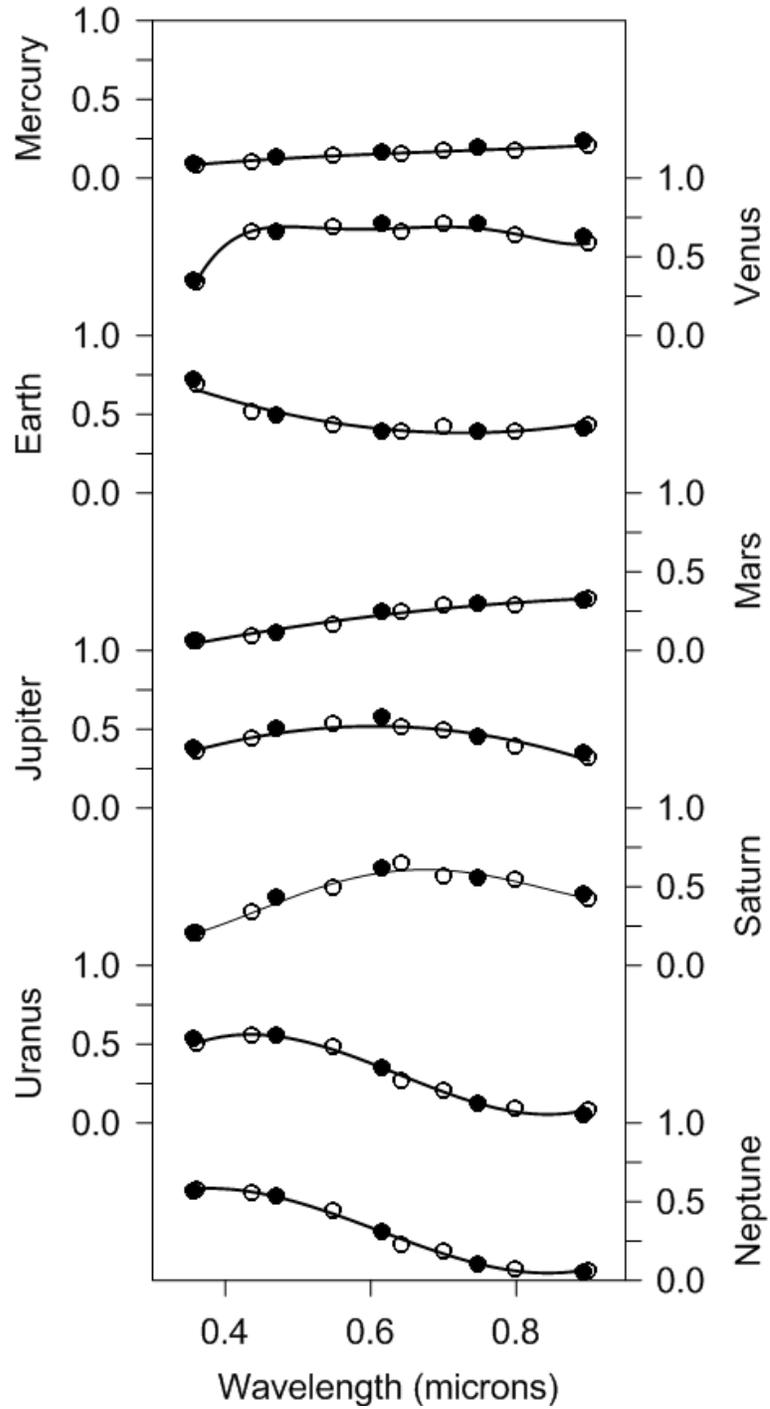

*Fig. 1. Albedos on the Sloan system (filled circles) and the Johnson-Cousins system (open circles) are consistent across wavelengths and planets. Note that the U symbols may overlap the u' symbols, and likewise with the I and z' symbols.*

7. Applications to exo-planets

The establishment of reference planetary magnitudes in our solar system allows for the estimation of apparent magnitudes for exo-planets. This information is useful for coronagraphic detection and for observation of such bodies when information about the signal-to-noise ratio is needed. Furthermore, an understanding of phase curve behaviors for different types of solar system planets can be used to characterize exo-planets even if the bodies do not transit their parent star. For example, a barren planet like Mercury will have a steep illumination phase curve, while a cloudy body like Venus will have a gentle curve and that of a partly cloudy orb like the Earth will be intermediate between the two.

The general guidelines summarized above were established in a paper (Mallama 2009) which has been cited in many exo-planet studies. Some examples of those investigations have included characterization of the potential super-Venus Kepler-69c (Kane et al. 2013), mapping of clouds on Kepler-7b (Garcia-Muñoz and Isaak, 2015) and the general approach to light scattering from exo-planet atmospheres (Zugger et al. 2010).

A particularly important phase angle is $90^o$ where exo-planets are near their greatest apparent separation from the central star. The u', g', r', i' and z' values of $M_1(90)$ are given for the terrestrial planets Mercury, Venus and Earth in the top portion of Table 8. Since the Sloan system is directly tied to energy units, magnitude values can be converted to fluxes and then combined arithmetically. Besides this convenience, the Sloan system was chosen over the Johnson-Cousins system for these computations because newer instruments generally contain Sloan filters. The column labeled $\Sigma$ gives a panchromatic magnitude which extends from about 0.3 to 1.0 µm.

The top portion of Table 8 also lists the equivalent values for Jupiter and Neptune. These two planets were chosen to represent gas giants and ice giants, respectively. Jupiter has a fairly smooth spectrum, at least at low resolution, which is not very dissimilar from Venus in its broad contour. By contrast, the red and near-IR portions of Neptune's spectrum are heavily blanketed by methane absorption bands, giving it a distinctly blue color. Since the illumination phase functions of these planets have only been determined over small phase angles, we took the magnitude corrections at $90^o$ to be the same as those for cloudy Venus.

Much of the challenge in observing exo-planets lies in distinguishing their relatively faint light in the presence of a much brighter star in the immediate proximity. Therefore, the middle portion of Table 8 indicates the delta magnitudes between the planets and the Sun for each Sloan band. The column Δmag lists the panchromatic delta stellar magnitude. The final two columns, $R_E$ and $R_J$, list those magnitudes for planets whose radii are equal to that of the Earth and to that of Jupiter, respectively.

The band-specific delta magnitudes (u', g', r', i' and z') would be practically the same for any exo-planet orbiting a star of any spectral type because the bodies shine by reflected light. However, the panchromatic delta magnitudes would differ measurably. A bluish planet like Neptune, for example, would appear brighter in relation to an early type (bluish) star and vice-versa for a reddish planet like Mars. Thus, the bottom portion of Table 8 indicates corrections to panchromatic differential magnitudes for each planet as a function of the stellar spectral type. The correction varies by more than one magnitude in the case of Neptune, from -0.50 for a B0 star to +0.58 for an M0 star. The corrections for a red planet like Mars (not in the Table) would trend in the opposite direction from +0.52 to -0.18.

Table 8. Magnitudes at phase angle $90^o$ for exo-planets

```
Absolute magnitudes for the planets of a G2 main sequence star at 1 AU
            u'      g'      r'      i'      z'      Σ
Mercury    4.34    2.57    1.84    1.55    1.40    0.23
Venus     -0.36   -2.57   -3.12   -3.24   -3.20   -4.59
Earth     -0.75   -1.77   -1.94   -2.07   -2.09   -3.57
Jupiter   -5.81   -7.57   -8.18   -8.05   -7.84   -9.48
Neptune   -4.00   -5.38   -5.25   -4.18   -3.43   -6.44

Delta magnitudes for planets of a G2 main sequence star
                                                -- panchromatic --
            u'      g'      r'      i'      z'    actual   R_E     R_J
Mercury   29.36   29.02   28.73   28.55   28.37   28.63   26.55   21.37
Venus     24.66   23.88   23.77   23.76   23.77   23.81   23.70   18.52
Earth     24.27   24.68   24.95   24.93   24.88   24.84   24.84   19.66
Jupiter   19.21   18.88   18.71   18.95   19.13   18.93   24.10   18.93
Neptune   21.02   21.07   21.64   22.82   23.54   21.96   24.89   19.72
```

```
Add to panchromatic delta magnitudes for other spectral types
  Sp. Type   B0       A0       F0       G0       K0       M0
Mercury     +0.33    +0.13    +0.05    +0.01    -0.03    -0.14
Venus       +0.26    +0.06    +0.03    +0.01    -0.02    -0.05
Earth       -0.15    -0.05    -0.02     0.00    +0.02    +0.04
Jupiter     +0.11    +0.01     0.00     0.00     0.00    +0.05
Neptune     -0.50    -0.30    -0.14    -0.01    +0.09    +0.58
```

For the convenience of those who prefer to work in fluxes and flux ratios we have added Table 9 which parallels Table 8. Fluxes are given in ergs/cm^2/s/Hz at the band centers according to Equation 5 where $F_v$ is the flux and *m* is the magnitude. (See Smith et al., 2002 for the equation and the definition of the AB system absolute monochromatic flux.) We also remind readers that the planetary albedos are listed in Table 7.

$$F_v = 10^{\{(48.60 + m)/-2.5\}}$$

Equation 5

Table 9. Fluxes at phase angle 90° for exo-planets

```
Absolute fluxes for the planets of a G2 main sequence star at 1 AU
            u'        g'        r'        i'        z'        Σ
Mercury   6.67E-22  3.40E-21  6.67E-21  8.71E-21  1.00E-20  2.95E-20
Venus     5.06E-20  3.88E-19  6.41E-19  7.16E-19  6.95E-19  2.49E-18
Earth     7.25E-20  1.86E-19  2.17E-19  2.45E-19  2.49E-19  9.69E-19
Jupiter   7.65E-18  3.88E-17  6.77E-17  6.01E-17  4.99E-17  2.24E-16
Neptune   1.44E-18  5.17E-18  4.56E-18  1.70E-18  8.59E-19  1.37E-17

Flux ratios for planets of a G2 main sequence star
                                                 ----- panchromatic -----
            u'        g'        r'        i'        z'      actual     R_E       R_J
Mercury   1.80E-12  2.47E-12  3.22E-12  3.80E-12  4.49E-12  3.53E-12  2.41E-11  2.84E-09
Venus     1.37E-10  2.81E-10  3.10E-10  3.12E-10  3.12E-10  2.99E-10  3.31E-10  3.90E-08
Earth     1.96E-10  1.34E-10  1.05E-10  1.07E-10  1.12E-10  1.16E-10  1.16E-10  1.37E-08
Jupiter   2.07E-08  2.81E-08  3.27E-08  2.62E-08  2.24E-08  2.69E-08  2.28E-10  2.69E-08
Neptune   3.91E-09  3.74E-09  2.20E-09  7.42E-10  3.85E-10  1.65E-09  1.10E-10  1.30E-08
```

```
Multiplier for panchromatic flux ratios for other spectral types
  Sp. Type    B0    A0    F0    G0    K0    M0
Mercury     0.74  0.89  0.95  0.99  1.03  1.14
Venus       0.78  0.94  0.97  0.99  1.02  1.05
Earth       1.15  1.04  1.02  1.00  0.98  0.96
Jupiter     0.90  1.00  1.00  1.00  1.00  0.96
Neptune     1.59  1.32  1.14  1.01  0.92  0.59
```

To conclude this section we discuss our exo-planet results in the context of related work. Dyudina et al. (2005) derived albedos and orbital light curves for exo-planets resembling Jupiter and Saturn. They used the observed anisotropic scattering functions derived by Tomasko et al. (1978) and Smith and Tomasko (1984) for Jupiter, and by Tomasko and Doose (1984) and Dones et al. (1993) for Saturn. Those derived functions were based on imaging data from the Pioneer 10 and 11 missions and they were originally fitted in order to characterize the atmospheres of the two planets according to their scattering properties. Dyudina et al. repurposed the parametric fits to compute the integrated planetary brightness as a function of the orbital phase angle. Since exo-planets are extremely distant from the Earth, their orbital phase angle is practically the same as the illumination phase angle. In fact, the two are identical at angle zero which corresponds to the geometric albedo. Thus, we compared our observed albedos with those modeled by Dyudina et al. as described below.

Figure 6 of Dyudina et al. shows their planetary light-curves for spherically modeled Jupiter and Saturn in red light (0.6 – 0.7 μm). Figure 7 includes the realistic case of a 10% oblate Saturn and shows the light-curve as seen from the equator and from 45 degrees latitude. There is no numerical data to accompany these plots and the case of a realistic 6% oblate Jupiter is not shown. However, we were able to estimate the values needed for the comparison by digitizing the figures and by approximating the effect of oblateness. Beginning with Jupiter, we found that the geometric albedo in Figure 6 is about 0.54. Now, judging from the geometric albedos for Saturn (discussed below) we estimate that 1% of flattening corresponds to an albedo reduction of approximately 1%. Thus, the geometric albedo of 6% oblate Jupiter would be about 0.51. For comparison with our study we choose the $R_C$ band because its effective wavelength, 0.641 μm, and full-width-at-half-maximum, 0.158 μm, are the closest fit to the red wavelength range cited above. According to Table 7 in this paper, the albedo of Jupiter in the $R_C$ band is 0.513 which validates the modeled value given above for oblate Jupiter.

For Saturn, the geometric albedo for the spherically modeled planet as seen from the equator is about 0.47 while that for the realistic 10% oblate planet is 0.43. (Thus, the 10% oblate planet is 10% dimmer which is the reason we applied a correction of 1% per degree of flattening for Jupiter above.) As for the comparison to our Saturn data, the geometric albedo for the $R_C$ band from Table 7 is 0.646 which is 50% brighter than the 10% oblate value read from the light-curve of Dyudina et al. and thus the difference requires an explanation. Part of the disagreement may be that our result is too bright. Saturn with its ring system is the most difficult planet for which to derive a heuristic brightness model because the inclination of the rings causes the overall luminosity to vary by more than a factor of two (Mallama, 2012). Therefore, the albedo of the globe of the planet itself could be in error, perhaps by 10% or more but certainly not by 50%. So, an additional explanation is still required. Dyudina et al. state that they averaged the scattering coefficients for the two types of atmospheric bands of Saturn, namely, the belts and zones. Simple averaging could lead to an overall albedo value that is too low because the zones actually cover a much larger range of latitudes than do the belts and because they are far brighter. Figure 3 of Mallama (2014) shows the predominance of the zones especially in the equatorial region of Saturn where they are preferentially projected into the equatorial view of the planet. So, the difference between the observed Saturnian albedo in this paper and that from Dyudina et al. may be explained if the former is too large and the latter is too small.

Only a restricted range of outer planet illumination phase angles may be observed from Earth, so phase curve comparisons are limited. For Jupiter and Saturn the maximum phase angles are only 12 and 6 degrees, respectively, whereas Dyudina et al. calculated the light curve (or phase curve) over all 360 degrees. While our observations cannot validate very much of the modeled outer planet phase curves, we note that there is good agreement for Jupiter at 12 degrees where both sources indicate about 6% dimming relative to angle zero. Saturn cannot be validated at all because the dimming at 6 degrees according to the model is only about 1%. Furthermore, our observed phase curve for Saturn is more sensitive to the large effect of the inclination of the ring than to the smaller effect of the illumination phase. While that parameterization is the best heuristic available for the purpose of representing Saturn's brightness for ephemerides it does not produce a very accurate illumination phase curve.

Before proceeding from Jupiter and Saturn to Uranus and Neptune, we mention that Dyudina and her colleagues built upon much of the work cited above in a later study (Dyudina et al. 2016). However, the data in the modeled light curves in their Figure 2 of the 2016 paper cannot be tested with our observations because their narrow and specialized band passes ($CH_4$ and an atmospheric window) and the UV band pass at 0.258 μm do not correspond with those that we studied.

Pollack et al. (1986) used data from the Voyager spacecraft as well as some ground-based observations to investigate the albedos of Uranus and Neptune. The minimum phase angles for Voyager used in that study were 43 degrees for Uranus and 25 degrees for Neptune. These are much larger than the angles that can be observed from the ground (3 degrees and 2 degrees, respectively), though, so they cannot be compared with our albedos. Those authors also considered results derived from ground-based spectrophotometry obtained by Neff et al. (1985). However, Karkoschka (1998) had already evaluated that work and noted some of its deficiencies when he published his more definitive spectrophotometry. Since we have incorporated Karkoschka's data in the results derived in the present paper we do not discuss the results from Pollack et al. and Neff et al. any further.

Following up on the work of Pollack et al., Rages et al. (1991) used Voyager data to analyze the scattering properties in two latitude bands of the troposphere and lower stratosphere of Uranus. However, the results were not integrated over the disk of Uranus and the minimum phase angle of the data was 14 degrees, so it is not useful to compare our data with those results.

Thus, we have shown that the currently available data for Uranus and Neptune cannot be validated with our results. We further point out that both planets are so highly variable in brightness that any such attempted comparison would be risky. Schmude et al. (2015) showed that Uranus exhibits large brightness variations especially at red and near-IR wavelengths. This is an example of an apparent brightness variation which, in this case, is due to the viewing geometry. When the methane-depleted polar regions of Uranus face toward the Sun and Earth, the planet appears brighter because there is less absorption in the methane bands which dominate the planet's spectrum especially from about 0.6 μm out into the IR. Conversely, Uranus appears fainter when the methane-laden equatorial latitudes are projected to the Sun and Earth. These latitude-dependent variations are characterized in Schmude et al. and also in the appendix to this paper. As exo-planet observations become more powerful and accurate, it seems likely that apparent brightness variations will be detected on those bodies as well, though possibly for other reasons.

Finally, we note that Neptune undergoes short-term rotational brightness variations as well as long-term intrinsic changes. Schmude et al. (2016) demonstrated that clouds rotating across the apparent disk of Neptune significantly alter its integrated brightness on a diurnal time-scale. Furthermore, a long-term brightening occurred from approximately the year 1980 until 2000. The increased brightness was about 10% or slightly more at blue and green wavelengths, while it was even larger in the red and near-IR portions of the spectrum. The cause of this long-term brightening is still not known. As with the apparent brightness variations of Uranus, intrinsic variability like Neptune will probably also be observed in exo-planets.

8. Application to Planet Nine

Trujillo and Sheppard (2014) pointed out the peculiar clustering of orbital elements among several transneptunian object (TNOs). Batygin and Brown (2016) then derived the orbital parameters of a proposed body called Planet Nine (P9) that could explain that clustering. They also estimated a mass of ~10 $M_E$ or greater. Subsequently, Linder and Mordasini (2016) modeled the evolution of P9 based on this mass and the assumption that it is similar to the ice giants Uranus and Neptune. They calculated a radius of 3.66 $R_E$ as well as temperatures of 66$^o$ K at 1 bar and 47$^o$ K at $\tau$ = 2/3 for the present epoch. They also estimated the V, $R_C$ and $I_C$ values of such a body.

The magnitudes of P9 estimated in this study are given in Tables 10 – 12 where the planets Neptune, Jupiter and Pluto, respectively, are used for brightness models as described in the following paragraphs. In each case though the radius of P9 is scaled to 3.66 $R_E$. Sloan magnitudes are listed in the tables because those filters are being installed in most new astronomical instruments. Additionally, we show Johnson-Cousins magnitudes for comparison with Linder and Mordasini and for completeness. The first line of each table lists the magnitudes at a distance of one AU from the Sun and Earth. Lines 2 and 3 indicate the apparent magnitudes of P9 at perihelion (280 AU) and aphelion (1120 AU), respectively. These distances are based on the semi-major axis (700 AU) and the eccentricity (0.6) of the orbit indicated by Batygin and Brown. Line 4 lists our magnitude estimates at a distance of 700 AU for comparison with those reported by Linder and Mordasini which are shown in line 5; the differences are given in line 6. The magnitudes in line 7 are at a distance of 624 AU which corresponds to a true anomaly (*v*) of 118$^o$. Fienga et al. (2016) determined that anomaly as the most likely position of P9 at the present time based upon their perturbation analysis.

The results given in Table 10 were derived from the reference values of Neptune indicated in Tables 3 and 5 of this paper. While the V magnitude agrees closely with that of Linder and Mordasini, the $R_C$ and $I_C$ values are fainter by 0.6 and 1.9 magnitudes, respectively. Linder and Mordasini assumed a constant geometric albedo of 0.41 regardless of wavelength. Meanwhile, our albedos for $R_C$ and $I_C$ from Table 7 are far smaller at 0.226 and 0.072, respectively. The actual faintness of Neptune in the red and near-IR wavelengths is due to its strong methane absorption bands as noted previously.

P9 might not be an ice giant like Neptune though. Furthermore, if its temperature is much lower than that indicated by Linder and Mordasini it might not support an observable methane atmosphere and its spectrum would be quite dissimilar from Neptune. Therefore, we also estimated the magnitudes of P9 based on those of Jupiter and of Pluto. Jupiter was chosen as an example of a gas giant lacking very strong methane absorption, while Pluto represents a true TNO.

The resulting magnitudes based on Jupiter, listed in Table 11, show a much closer agreement with Linder and Mordasini in the $R_C$ and $I_C$ bands. Our value of $R_C$ is 0.3 magnitude brighter and the two measures of $I_C$ are essentially equal.

We did not establish reference magnitude for Pluto in this paper and a search of the literature did not locate values for $R_C$ and $I_C$. Therefore we synthesized magnitudes from the spectral profile shown in Figure 3 of Lorenzi et al. (2016). Since their data are given as a relative reflectance value rather than in absolute energy units, we adjusted our synthetic magnitudes such that the albedo corresponding to our V value agreed with the average albedo cited in their Table 1. A small extrapolation beyond their limits of 0.40–0.93 µm was also required. The resulting V, $R_C$ and $I_C$, magnitudes (not affected by extrapolation) which are listed in Table 12 differ by no more than 0.4 from those of Linder and Mordasini.

When the magnitudes of P9 based on Jupiter, Neptune and Pluto are examined together they lead to a somewhat unexpected result. Specifically, the greatest RMS difference for the V, $R_C$ and $I_C$ bands compared to Linder and Mordasini is for the Neptune-based magnitudes (RMS = 0.13) whereas their model is for an planet like Uranus or Neptune. Meanwhile, the corresponding RMS differences for Jupiter and Pluto are much smaller (0.02 and 0.04, respectively). So, we find magnitudes that agree most closely with Linder and Mordasini for the two bodies which they did not model.

Finally, our analysis has implications for telescopic searches aimed at locating P9. In particular, if the body is an ice giant at a temperature warm enough to support an atmosphere with abundant methane, like Neptune, it will be especially faint at red and near-IR wavelengths. Therefore, searches at wavelengths short-ward of about 0.6 µm, where P9 would be much brighter, could be advantageous. Recording data in the Sloan g' filter or the Johnson-Cousins V filter would also minimize interference from faint field stars; those background objects are disproportionately red due to the higher number density of cool stars and to the effect of interstellar reddening.

Table 10. Magnitude for P9 referenced to Neptune

```
AU       U     B     V     R     I     R_C   I_C   u'    g'    r'    i'    z'
   1  -6.26 -6.43 -6.82 -6.39 -5.65 -6.49 -5.59 -5.36 -6.73 -6.57 -5.48 -4.62
 280  18.21 18.04 17.65 18.08 18.82 17.98 18.88 19.11 17.74 17.90 18.99 19.85
1120  24.23 24.06 23.67 24.10 24.84 24.00 24.90 25.13 23.76 23.92 25.01 25.87
 700 (This paper) 21.63             21.96 22.86
 700 (L & M)      21.7              21.4  21.0
 700 (delta)      -0.1              +0.6  +1.9
 624  21.69 21.52 21.13 21.56 22.30 21.46 22.36 22.59 21.22 21.38 22.47 23.33
```

Table 11. Magnitude for P9 referenced to Jupiter

```
AU       U     B     V     R     I     R_C   I_C   u'    g'    r'    i'    z'
   1  -5.75 -6.18 -7.04 -7.49 -7.36 -7.39 -7.43 -4.93 -6.68 -7.26 -7.11 -6.79
 280  18.72 18.29 17.43 16.98 17.11 17.08 17.04 19.54 17.79 17.21 17.36 17.68
1120  24.74 24.31 23.45 23.00 23.13 23.10 23.06 25.56 23.81 23.23 23.38 23.70
 700 (This paper) 21.41             21.06 21.02
 700 (L & M)      21.7              21.4  21.0
 700 (delta)      -0.3              -0.3  0.0
 624  22.20 21.77 20.91 20.46 20.59 20.56 20.52 23.02 21.27 20.69 20.84 21.16
```

Table 12. Magnitude for P9 referenced to Pluto

```
AU       U     B     V     R     I     R_C   I_C   u'    g'    r'    i'    z'
   1  -5.38 -5.93 -6.96 -7.72 -8.12 -7.50 -7.87 -4.44 -6.45 -7.26 -7.48 -7.48
 280  19.09 18.54 17.51 16.75 16.35 16.97 16.60 20.04 18.02 17.22 16.99 16.99
1120  25.11 24.57 23.53 22.78 22.37 22.99 22.63 26.06 24.04 23.24 23.01 23.01
 700 (This paper) 21.49             20.95 20.58
 700 (L & M)      21.7              21.4  21.0
 700 (delta)      -0.2              -0.4  -0.4
 624  22.57 22.02 20.99 20.23 19.83 20.45 20.08 23.52 21.50 20.70 20.47 20.47
```

9. Summary and Conclusions

This paper provides complete listings of planetary magnitudes and albedos in the two principal wide-band photometry regimes in use today, namely, the Sloan system and the Johnson-Cousins system. The Johnson-Cousins magnitudes are excerpted from modern photometric studies of the planets and augmented with synthetic magnitudes derived from published spectrophotometry. The Sloan magnitudes are an average of photometric and synthetic measures as well as the results of transformations from Johnson-Cousins values. Albedos derived from magnitudes on two systems are consistent across all 8 planets and all 12 band-passes. This agreement validates the albedos and magnitudes on both photometric systems.

Applications of the Sloan results to exo-planet research are discussed. These include estimation of signal-to-noise ratios for coronoagraphic exo-planet searches in the visible and near-visible portions of the spectrum, as well as characterization of such bodies based upon their illumination phase curves.

The Johnson-Cousins and Sloan magnitudes of the proposed Planet Nine are also estimated based upon the magnitudes of Jupiter, Neptune and Pluto. Our estimates are compared with those of Linder and Mordasini (2016) and we note the implications for telescopic searches for P9.

The appendix provides detailed information about individual planets in separate sections devoted to the eight bodies. Each section outlines the procedures employed for obtaining new Sloan photometry, lists those observational results, describes the illumination phase function and other phase effects that determine the instantaneous magnitude of the planet, and recounts the method by which synthetic magnitudes were derived from published spectrophotometry.

Acknowledgments

We wish to thank the American Astronomical Society for a Small Research Grant which funded some of the hardware used to acquire data for this study. We also appreciate several helpful discussions with R.E. Baker regarding the extraction of photometric data from CCD images, communications with D.R. Skillman regarding exo-planets and Planet Nine, correspondence with S.S. Limaye and D.R. Williams regarding the albedo of Venus, and dialogs with A. Pickles and E. Mamajek regarding the Sloan magnitudes for main sequence stars. R. West and an anonymous reviewer offered suggestions that led to an improved paper. H.M. Aelion helped with the revision.

Tomasko, M.G. and Doose, L.R. 1984.Polarimetry and photometry of Saturn from Pioneer 11: Observations and constraints on the distribution and properties of cloud and aerosol properties. Icarus 58, 1-34 (1984) . http://www.sciencedirect.com/science/article/pii/0019103584900964.

Trujillo, C. A., & Sheppard, S. S. 2014. A Sedna-like body with a perihelion of 80 astronomical units. Nature, 507, 471-474. http://www.nature.com/nature/journal/v507/n7493/full/nature13156.html.

Vincendon, M., Langevin, Y., Poulet, F., Pommerol, A., Wolff, M., Bibring, J.-P., Gondet, B. and Jouglet, D. 2009. Yearly and seasonal variations of low albedo surfaces on Mars in the OMEGA/MEx dataset: Constraints on aerosols properties and dust deposits, Icarus, 200, 395-405. https://arxiv.org/ftp/arxiv/papers/1103/1103.3426.pdf.

Warell, J. 2004. Properties of the Hermean regolith: IV. Photometric parameters of Mercury and the Moon contrasted with Hapke modeling. Icarus, 167, 271-286. http://www.sciencedirect.com/science/article/pii/S0019103503003543.

Zugger, M.E., Kasting, J.F., Williams, D.F., Kane, T.J. and Philbrick, C.R. 2010. Light scattering from exoplanet oceans and atmospheres. Ap.J. 723, 1168-1179. https://arxiv.org/ftp/arxiv/papers/1006/1006.3525.pdf.
34

Appendix

The body of this paper discusses the eight planets as a set. However, each one is unique and the availability of observed, synthetic and transformed data is not uniform. The sections in this appendix address the planets individually. Each section begins with the procedures used in obtaining Sloan photometry and lists the $M_1(\alpha)$ magnitudes with the corresponding Modified Julian Day (MJD) of observation. The determination of $M_1(0)$ from $M_1(\alpha)$ requires the use of phase functions pertaining to any given planet. The illumination coefficients and other values of these functions were determined in the Johnson-Cousins photometric system in previous studies. Thus, it was necessary to derive their corresponding values for the Sloan system by interpolation. The observed constants on the Johnson-Cousins system and the interpolated Sloan constants for the phase functions pertaining to each planet are, therefore, also provided along with the equations for computing $M_1(0)$ magnitudes. $M_1(0)$ corresponds to the zero order coefficient, $C_0$, of each phase function. For the Sloan magnitudes, this coefficient was initially set to zero and was then solved for based on observational data. The arc units of the phase functions are degrees. The sources of spectrophotometric data and the methods of deriving synthetic magnitudes are also discussed.

A-1. Mercury

Sloan system photometry of Mercury was obtained during daylight, when the planet was well above the horizon, in order to minimize the effects of atmospheric extinction. The telescope aperture was reduced to 25 mm diameter to prevent saturation of the CCD images by the bright sky background. A measurement aperture of about 15 arc seconds radius was used to extract raw magnitudes from the images. The observations of Mercury and the comparison star, HD 34029, recorded in April and May 2015 are listed in Table A-1.1.

Table A-1.1 Sloan $M_1(\alpha)$ magnitudes of Mercury

| MJD | u' | +/- | g' | +/- | r' | +/- | i' | +/- | z' | +/- |
|---|---|---|---|---|---|---|---|---|---|---|
| 57138.66 | 3.566 | 0.116 | 1.717 | 0.052 | 1.002 | 0.032 | 0.697 | 0.043 | 0.572 | 0.035 |
| 57138.78 | 3.675 | 0.046 | 1.754 | 0.038 | 1.048 | 0.017 | 0.752 | 0.031 | 0.578 | 0.020 |
| 57144.86 | 3.992 | 0.073 | 2.341 | 0.062 | 1.745 | 0.022 | 1.439 | 0.027 | 1.309 | 0.032 |

The illumination phase function for Mercury in the Johnson-Cousins system was determined from observations made with the Solar and Heliospheric Observatory satellite and from the ground by Mallama et al. (2002). The observed phase angles ranged from 2 to 170 degrees. While coefficients were only derived for the V band, it is reasonable to assume that the function is fairly constant with wavelength since Mercury is grey and airless. Therefore the V band phase coefficients listed in A-1.2 and the polynomial in Equation A-1.1 were applied to all Sloan bands when solving for $M_1(0)$.

$$M_1(\alpha) = \sum_{i=1}^{n} C_i\ \alpha^i$$

Equation A-1.1

Table A-1.2. Illumination phase function coefficients for Mercury

| C# | Coefficient |
|---|---|
| 0 | −0.694 |
| 1 | +6.617E−02 |
| 2 | −1.867E−03 |
| 3 | +4.103E−05 |
| 4 | −4.583E−07 |
| 5 | +2.643E−09 |
| 6 | −7.012E−12 |
| 7 | +6.592E−15 |

Our geophysical interpretation of the data in Mallama et al (2002) has been cited in studies of Mercury by Warell (2004) and many later investigators. Likewise, the data have been used to study the general topic of opposition phenomena exhibited by solar system objects by Rosenbush et al. (200). Finally, Hilton (2005) analyzed the V-band data in his derivation of the magnitude model which is currently used in the Astronomical Almanac.

Synthetic magnitudes were derived from the study by Izenberg et al. (2014) which was based on data from the MASCS instrument on the MESSENGER spacecraft. The reflectance values digitized from the spectrophotometry in Figure 5 of that paper were converted to geometric albedos by applying a factor of 2.00. The conversion constant was obtained by noting that their radiance factor of 0.071 at 0.564 μm is exactly half of the V-band (0.550 μm) geometric albedo reported by Mallama et al. The geometric albedos were then converted to energy units by factoring in the solar flux and the planet's radius.

A-2. Venus

Like Mercury, the Sloan system photometry of Venus was obtained during daylight when the planet was well above the horizon in order to minimize the effects of atmospheric extinction. The telescope aperture was reduced to 25 mm diameter in order to prevent saturation of the CCD images by the bright sky background. Even at that aperture, though, the extreme intensity of Venus required defocusing of the planet in order to avoid saturation of the images. On the other hand, the fainter comparison star, HD 34029, had to be well focused in order to provide a high signal-to-noise ratio. Therefore measurement apertures of approximately 25 and 50 arc seconds radius were used to measure raw magnitudes from images of the comparison and the planet, respectively. The observations in Table A-2.1 were recorded in March and April 2015.

Table A-2.1 Sloan $M_1$ (α) magnitudes of Venus

```
    MJD        u'    +/-      g'    +/-      r'    +/-      i'    +/-     z'    +/-
 57111.75  -1.236 0.178   -3.249 0.092   -3.862 0.029   -3.965 0.032  -3.939 0.032
 57111.86  -0.986 0.063   -3.235 0.035   -3.877 0.021   -4.003 0.018  -4.053 0.023
 57114.77  -0.918 0.166   -3.350 0.039   -3.865 0.026   -4.006 0.034  -3.983 0.018
 57114.85  -1.202 0.296   -3.383 0.038   -3.855 0.015   -3.969 0.027  -3.999 0.016
```

The illumination phase function for Venus was determined in the Johnson-Cousins B, V, R and I bands by Mallama et al. (2006) from SOHO and ground-based observations. The observed phase angles ranged from 2 to 179 degrees. The U band coefficients were set equal to those of the B band since there are no strong color variations with phase angle evident at the other wavelengths. The phase function up to 165 degrees is well characterized by the observed Johnson-Cousins coefficients listed in Table A-2.2. The values interpolated for the Sloan system are in Table A-2.3. The I-band filter used by Mallama et al. did not match the transmission curve of the standard Johnson-Cousins band-pass very closely and, therefore, that magnitude is rather uncertain.

Table A-2.2. Observed Johnson-Cousins illumination phase function coefficients for Venus

```
C#          U              B              V              R              I
0     -3.178E+00     -3.678E+00     -4.384E+00     -4.947E+00     -5.076E+00
1      5.650E-03      5.650E-03     -1.044E-03      5.853E-03     -7.863E-03
2      2.438E-04      2.438E-04      3.687E-04      1.475E-04      4.571E-04
3     -2.001E-06     -2.001E-06     -2.814E-06     -4.645E-07     -3.024E-06
4      6.952E-09      6.952E-09      8.938E-09      1.220E-09      7.766E-09
```

Table A-2.3. Interpolated Sloan illumination phase function coefficients for Venus

```
C#          u'             g'             r'             i'             z'
0      0.000          0.000          0.000          0.000          0.000
1      5.650E-03      3.909E-03      2.014E-03      2.554E-03     -7.390E-03
2      2.438E-04      2.763E-04      2.706E-04      2.219E-04      4.464E-04
3     -2.001E-06     -2.213E-06     -1.772E-06     -1.080E-06     -2.936E-06
4      6.952E-09      7.468E-09      5.516E-09      2.795E-09      7.540E-09
```

Muñoz et al. (2014) used the phase curve from Mallama et al. (2006) to investigate the optical glory in the atmosphere of Venus. Bailey (2007) cited the same phase curve as one method to study the atmospheres of exo-planets. Hilton (2005) used the V data in his derivation of the magnitude model which is currently used in Astronomical Almanac.

As with Mercury, data from the MASCS instrument on the MESSENGER spacecraft was used to synthesize magnitudes. The study by Perez-Hoyos et al. (2013) indicates a reflectivity factor of 0.532 at 0.564 μm while Mallama et al. list the geometric albedo 0.67 for the Johnson-Cousins V band at 550 μm. Thus, the ratio 1.26 was applied to the values which were digitized from the spectrophotometry in Figure 2 of Perez-Hoyos et al. The geometric albedos were then converted to energy units by taking account of the solar flux and the planet's radius. The synthetic magnitudes for the I and z' bands are significantly brighter than the corresponding photometric values and, therefore, have not been included in the analysis. These difference may be due to the non-zero phase angle of the spectral observations.

A-3. Earth

Spectrophotometric data for the Earth were derived from energy values measured by the Extrasolar Planet Observation and Deep Impact Extended Investigation spacecraft listed in Table 2 of Livengood et al. (2011). The illumination phase curve was a spline fit to the synthetic phase curve of the Earth for the 'realistic clouds' case plotted in Figure 7 of Tinetti et al. (2006). Representative values of the phase function, normalized to zero magnitudes at phase angle zero, are listed in Table A-3.1.

Table A-3.1. Illumination phase function for the Earth

```
    Angle   Function
        0      0.000
       45      1.123
       90      2.069
      135      3.801
```

Since the EPOXI data consisted of only 7 flux values which did not quite extend over the entire response curve of all the band-passes, interpolation and extrapolation were applied. The long wavelength values were determined by performing a linear extrapolation from the EPOXI 0.85 and 0.95 μm bands. The short wavelength values could not be linearly extrapolated because the UV flux distribution is not linear. Therefore an average of the normalized UV flux distributions of Venus (a cloudy planet) and Mars (a mostly clear planet) was used to extrapolate beyond the EPOXI value at 0.35 μm. The phase curve data of Tinetti et al. are for wavelengths from 0.5 to 0.9 μm, so there could be errors in the present study outside of those wavelengths.

A-4. Mars

For Sloan system photometry of Mars, the telescope aperture was reduced to 25 mm diameter. The planet and the comparison star, HD 124897, were defocused to approximately the same size on the CCD chip. Raw magnitudes were then extracted with a measurement aperture of about 40 arc seconds radius. The observations recorded in April and May 2014 are listed in Table A-4.1.

Table A-4.1 Sloan $M_1(\alpha)$ magnitudes of Mars

```
    MJD        u'    +/-      g'    +/-      r'    +/-      i'    +/-      z'    +/-
 56765.16    1.311 0.083   -0.759 0.037   -1.960 0.037   -2.291 0.032   -2.309 0.044
 56765.19    1.260 0.086   -0.782 0.040   -1.965 0.040   -2.314 0.036   -2.333 0.049
 56767.16    1.250 0.096   -0.695 0.038   -1.925 0.038   -2.246 0.032   -2.275 0.043
 56772.13    1.522 0.081   -0.540 0.049   -1.911 0.048   -2.256 0.029   -2.267 0.039
 56775.13    1.474 0.090   -0.535 0.045   -1.864 0.044   -2.234 0.028   -2.219 0.041
 56794.08    1.715 0.081   -0.356 0.051   -1.785 0.051   -2.234 0.029   -2.274 0.038
 56796.08    1.695 0.080   -0.351 0.046   -1.736 0.047   -2.142 0.028   -2.175 0.038
```

The globe of Mars exhibits distinctive albedo features. There are wide expanses of high reflectance terrain with a strong reddish coloration, as well as large areas characterized by lower reflectance and a more neutral color. These markings rotate across the visible disk each martian day causing the integrated brightness to change significantly. A variation of about 0.3 magnitude was observed in the Johnson-Cousins I-band by Mallama et al. (2007).

The orbital longitude of Mars also has a significant impact on its brightness. There is a geometrical factor caused by the variation of the sub-Earth and sub-Sun latitudes as well as a geophysical factor due to seasonal changes of the albedo features. The longitude phase curve causes the brightness of Mars to vary by roughly 0.2 magnitudes according to the Mallama et al.

The equation for the absolute magnitude from Mallama et al. is

$$M_1(\alpha, L_1, L_2) = C_0 + C_1 \alpha + C_2 \alpha^2 + L_1(\lambda_1) + L_2(\lambda_2)$$

Equation A-4.1

where $\alpha$ is the illumination phase angle, $\lambda_1$ is the rotation angle measured westward and $\lambda_2$ is the orbital longitude. The observed illumination phase angles ranged from a few degrees up to about 50 degrees. The coefficients of the illumination, rotation and longitude phase functions as observed on the Johnson-Cousins system and their interpolated values on the Sloan system are listed in Tables A-4.2 through A-4.7.

Table A-4.2. Observed Johnson-Cousins illumination phase function coefficients for Mars

| C# | U | B | V | R | I |
|---|---|---|---|---|---|
| 0 | 0.394 | -0.243 | -1.601 | -2.714 | -3.199 |
| 1 | 0.01680 | 0.02445 | 0.02267 | 0.01906 | 0.02205 |
| 2 | 0.0000228 | -0.0001131 | -0.0001302 | -0.0000956 | -0.0001931 |

Table A-4.3. Interpolated Sloan illumination phase function coefficients for Mars

| C# | u' | g' | r' | i' | z' |
|---|---|---|---|---|---|
| 0 | 0.000 | 0.000 | 0.000 | 0.000 | 0.000 |
| 1 | 0.01633 | 0.02399 | 0.02107 | 0.01978 | 0.02195 |
| 2 | 0.0000311 | -0.0001175 | -0.0001149 | -0.0001190 | -0.0001897 |

Table A-4.4. Observed Johnson-Cousins rotational phase function coefficients for Mars (selected values)

| $L_1$ (W) | U | B | V | R | I |
|---|---|---|---|---|---|
| 0 | 0.019 | 0.004 | 0.036 | 0.056 | 0.077 |
| 100 | 0.015 | -0.025 | -0.044 | -0.062 | -0.088 |
| 200 | 0.050 | -0.003 | -0.038 | -0.060 | -0.076 |
| 300 | 0.041 | 0.038 | 0.037 | 0.056 | 0.073 |

Table A-4.5. Interpolated Sloan rotational phase function coefficients for Mars (selected values)

| $L_1$ (W) | u' | g' | r' | i' | z' |
|---|---|---|---|---|---|
| 0 | 0.020 | 0.012 | 0.045 | 0.061 | 0.076 |
| 100 | 0.017 | -0.030 | -0.052 | -0.068 | -0.087 |
| 200 | 0.053 | -0.012 | -0.048 | -0.064 | -0.075 |
| 300 | 0.041 | 0.038 | 0.045 | 0.060 | 0.072 |

Table A-4.6. Observed Johnson-Cousins orbital phase function coefficients for Mars (selected values)

| $L_2$ | U | B | V | R | I |
|---|---|---|---|---|---|
| 0 | 0.000 | -0.038 | -0.029 | 0.014 | -0.006 |
| 100 | 0.000 | -0.031 | -0.024 | -0.075 | -0.074 |
| 200 | 0.000 | -0.075 | -0.016 | 0.031 | 0.047 |
| 300 | 0.000 | -0.020 | 0.006 | 0.053 | 0.080 |

Table A-4.7. Interpolated Sloan orbital phase function coefficients for Mars (selected values)

| $L_2$ | u' | g' | r' | i' | z' |
|---|---|---|---|---|---|
| 0 | 0.000 | -0.036 | -0.010 | 0.009 | -0.005 |
| 100 | 0.000 | -0.029 | -0.047 | -0.075 | -0.074 |
| 200 | 0.000 | -0.060 | 0.005 | 0.035 | 0.046 |
| 300 | 0.000 | -0.013 | 0.027 | 0.059 | 0.079 |

Mallama et al. (2007) also determined that global dust storms can produce a 15% brightness excess. This paper was cited by Vincendon et al. (2009) in their analysis of seasonal and yearly albedo variations of the planet based on data from the OMEGA infrared instrument on the Mars Express spacecraft.

Fluxes used in the computation of synthesized magnitudes were derived from the spectrophotometry of McCord and Adams (1969). Their figure 3, plots the geometric albedo of the 'integral disk (mostly bright regions)' for 0.3 to 1.2 µm averaged over several different martian seasons. No correction for phase was needed as these albedo measurements are geometric. Energy units were derived from albedo values based upon on the angular size of Mars and the solar flux.

A-5. Jupiter

The telescope aperture was reduced to 25 mm diameter for the Sloan system photometry of Jupiter. The planet and the comparison star, HD 34029, were both in focus on the CCD chip. Raw magnitudes were extracted from the CCD images with measurement apertures of about 18 arc seconds radius for the comparison and 40 arc seconds for Jupiter. The observations recorded in April and May 2014 are listed in Table A-5.1.

Table A-5.1 Sloan $M_1(\alpha)$ magnitudes of Jupiter

| MJD | u' | +/- | g' | +/- | r' | +/- | i' | +/- | z' | +/- |
|---|---|---|---|---|---|---|---|---|---|---|
| 57081.13 | -7.251 | 0.007 | -9.067 | 0.006 | -9.625 | 0.008 | -9.579 | 0.005 | -9.200 | 0.009 |
| 57082.11 | -7.268 | 0.010 | -9.101 | 0.005 | -9.653 | 0.004 | -9.583 | 0.005 | -9.231 | 0.011 |
| 57091.08 | -7.254 | 0.009 | -9.068 | 0.007 | -9.646 | 0.005 | -9.594 | 0.005 | -9.234 | 0.011 |

The jovian luminosity model is a simple quadratic illumination phase function as indicated in Equation A-5.1

$$M_1(\alpha, L_1, L_2) = C_0 + C_1 \alpha + C_2 \alpha^2$$

Equation A-5.1

where α is the illumination phase angle. The observed phase angles ranged from 0 to 12 degrees. Mallama and Schmude (2012) found brightness changes which correlate with variations in the atmospheric state but their amplitudes were just a few hundredths of a magnitude. The coefficients of the illumination phase function as observed on the Johnson-Cousins system and their interpolated values on the Sloan system are listed in Tables A-5.2 and A-5.3, respectively.

Table A-5.2. Observed Johnson-Cousins illumination phase function coefficients for Jupiter

| C# | U | B | V | R | I |
|---|---|---|---|---|---|
| 0 | -8.1060 | -8.5400 | -9.3950 | -9.8540 | -9.7180 |
| 1 | 0.00160 | -0.00096 | -0.00037 | -0.00086 | -0.00002 |
| 2 | 0.0012630 | 0.0006850 | 0.0006160 | 0.0004830 | 0.0003610 |

Table A-5.3. Interpolated Sloan illumination phase function coefficients for Jupiter

| C# | u' | g' | r' | i' | z' |
|---|---|---|---|---|---|
| 0 | 0.0000 | 0.0000 | 0.0000 | 0.0000 | 0.0000 |
| 1 | 0.00176 | -0.00080 | -0.00059 | -0.00066 | -0.00005 |
| 2 | 0.0012984 | 0.0006671 | 0.0005570 | 0.0004537 | 0.0003652 |

Fluxes used in the computation of synthesized magnitudes were derived from the spectrophotometry of Karkoschka (1998). Magnitudes were corrected for the 6.8 degree phase angle of Jupiter at the time of observation using Eq. A-5.1 and the coefficients from Tables A-5.2 and A-5.3.

A-6. Saturn

The telescope aperture was reduced to 50 mm diameter for the Sloan system photometry of Saturn. The planet and the comparison star, HD 138905, were both in focus on the CCD chip. Raw magnitudes were extracted from the images with measurement apertures of about 25 arc seconds radius for the comparison and 40 arc seconds for Saturn. The observations recorded in June 2014 are listed in Table A-6.1.

Table A-6.1. Sloan $M_1$ (α) magnitudes of Saturn

```
   MJD         u'    +/-       g'    +/-       r'    +/-       i'    +/-       z'    +/-
 56830.09   -6.828 0.031   -8.959 0.008   -9.754 0.005   -9.851 0.019   -9.696 0.009
 56831.10   -6.855 0.017   -8.946 0.014   -9.773 0.012   -9.848 0.020   -9.704 0.009
 56835.10   -6.834 0.023   -8.930 0.006   -9.759 0.013   -9.852 0.019   -9.677 0.009
 56836.09   -6.814 0.026   -8.945 0.008   -9.786 0.011   -9.880 0.019   -9.688 0.010
```

The inclination of Saturn's ring system imparts brightness changes that exceed those due to the illumination phase angle. The equation for absolute magnitude from Mallama (2012) is

$$M_1(\alpha,\beta) = C_0 + C_1 \sin(\beta) + C_2 \alpha - C_3 \sin(\beta) e^{(C_4 \alpha)}$$

Equation A-6.1

where α is the illumination phase angle and β is the inclination angle of the ring system. The observed phase angles ranged from 0 to 6 degrees while the inclination varied from 0 to 27 degrees. The coefficients as observed on the Johnson-Cousins system are listed in Table A-6.2 while the interpolated values on the Sloan system are listed in Table A-6.3.

Table A-6.2. Observed Johnson-Cousins system coefficients for Saturn

```
        C#        U           B           V           R           I
         0     -7.080      -7.842      -8.914      -9.587      -9.606
         1     -2.596      -2.002      -1.825      -1.865      -2.367
         2      0.020       0.034       0.026       0.032       0.032
         3      0.970       0.525       0.378       0.293       0.393
         4     -1.39       -2.33       -2.25       -3.07       -3.11
```

Table A-6.3. Interpolated Sloan system coefficients for Saturn

| C# | u' | g' | r' | i' | z' |
|---|---|---|---|---|---|
| 0 | 0.000 | 0.000 | 0.000 | 0.000 | 0.000 |
| 1 | -2.632 | -1.956 | -1.843 | -1.986 | -2.350 |
| 2 | 0.019 | 0.032 | 0.029 | 0.032 | 0.032 |
| 3 | 0.997 | 0.487 | 0.340 | 0.317 | 0.390 |
| 4 | -1.330 | -2.310 | -2.610 | -3.080 | -3.110 |

As with Jupiter, fluxes used in the computation of synthesized magnitudes were derived from the spectrophotometry of Karkoschka (1998). The Earth and Sun were very close to the plane of Saturn's rings at the time of the observation, so the inclination terms in Equation A-6.1 which include the coefficients $C_1$ and $C_3$ are practically zero. However, Saturn was not near opposition from the Sun, so the magnitudes were corrected for its 5.7 degree phase angle using coefficients $C_2$ from the Tables.

A-7. Uranus

The telescope was used at full aperture for the Sloan system photometry of Uranus and the comparison star HD 4628, and both were in focus on the CCD chip. Raw magnitudes were extracted from the images with measurement apertures of about 12 arc seconds radius. The observations recorded from October 2014 through November 2015 are listed in Table A-7.1.

Table A-7.1 Sloan $M_1$ (α) magnitudes of Uranus

| MJD | u' | +/- | g' | +/- | r' | +/- | i' | +/- | z' | +/- |
|---|---|---|---|---|---|---|---|---|---|---|
| 56943.17 | -5.590 | 0.055 | -6.991 | 0.040 | -6.893 | 0.039 | -5.875 | 0.060 | -4.984 | 0.031 |
| 57018.07 | -5.579 | 0.035 | -6.982 | 0.017 | -6.892 | 0.018 | -5.894 | 0.022 | -4.953 | 0.035 |
| 57036.01 | -5.572 | 0.033 | -6.986 | 0.016 | -6.900 | 0.016 | -5.879 | 0.020 | -4.998 | 0.032 |
| 57343.07 | -5.587 | 0.034 | -6.990 | 0.013 | -6.910 | 0.016 | -5.924 | 0.016 | -4.977 | 0.030 |
| 57347.10 | -5.579 | 0.032 | -6.990 | 0.012 | -6.904 | 0.012 | -5.916 | 0.016 | -5.004 | 0.031 |
| 57349.09 | -5.576 | 0.033 | -6.976 | 0.012 | -6.904 | 0.012 | -5.911 | 0.016 | -5.020 | 0.028 |

The maximum phase angle for Uranus is only 3 degrees, so its effect on the planet's brightness is negligible. However, the magnitude varies significantly with the latitude presented to the observer and the Sun especially at long wavelengths. This variation appears to be due to the presence of strong methane bands which vary in intensity with latitude. The strengths of those bands is determined, in turn, by the equator-to-pole gradient of methane abundance. The latitude-dependent Equation A-7.1 from Schmude et al. (2015) is

$$M_1(\beta) = C_0 + C_1 \beta$$

Equation A-7.1

where $\beta$ is the average of the absolute values of the sub-Earth and sub-Sun planetographic latitudes. Thus, $M_1(0)$ corresponds to Uranus viewed at sub-latitudes zero. The observed coefficients on the Johnson-Cousins system are listed in Table A-7.2 while the interpolated coefficients for the Sloan system are in Table A-7.3.

Table A-7.2. Observed Johnson-Cousins latitude coefficients for Uranus

| C# | U | B | V | R | I |
| --- | --- | --- | --- | --- | --- |
| 0 | −6.280 | −6.610 | −7.110 | −6.690 | −5.570 |
| 1 | −0.00048 | −0.00048 | −0.00084 | −0.00533 | −0.00279 |

Table A-7.3. Interpolated Sloan latitude coefficients for Uranus

| C# | u' | g' | r' | i' | z' |
| --- | --- | --- | --- | --- | --- |
| 0 | 0.0000 | 0.0000 | 0.0000 | 0.0000 | 0.0000 |
| 1 | −0.00047 | −0.00056 | −0.00283 | −0.00472 | −0.00287 |

Photometry in the near-IR may be problematic due to the methane bands which heavily blanket that portion of the Uranian spectrum causing a very uneven flux distribution. Some of the observations obtained in the study by Schmude et al. (2015) were made with an I filter that was poorly matched to the Johnson-Cousins I-band response curve. This mismatch introduced uncertainty into the transformation to the standard Johnson-Cousins system. Therefore, the photometric I magnitude was omitted from Table 3 and subsequent analysis. The transformation from Johnson-Cousins R and I to Sloan z' is, likewise, subject to considerable uncertainty. So, Table 5 does not contain a transformed Sloan magnitude. The $I_C$ and z' filters were well matched to the standard response curves and, therefore, their transformation uncertainties are small.

Like Jupiter and Saturn, fluxes used in the computation of synthesized magnitudes were derived from the spectrophotometry of Karkoschka (1998). The resulting magnitudes were corrected for the average absolute sub-latitude of Uranus, 49 degrees when the spectrophotometry was acquired.

A-8. Neptune

The telescope was used at full aperture for the Sloan system photometry of Neptune and the comparison star HD 213464, and both were in focus on the CCD chip. Raw magnitudes were extracted from the images with measurement apertures of about 12 arc seconds radius. The observations recorded from September 2014 through November 2015 are listed in Table A-8.1.

Table A-8.1 Sloan $M_1$ (α) magnitudes of Neptune

| MJD | u' | +/- | g' | +/- | r' | +/- | i' | +/- | z' | +/- |
|---|---|---|---|---|---|---|---|---|---|---|
| 56924.13 | -5.478 | 0.019 | -6.936 | 0.020 | -6.703 | 0.019 | -5.760 | 0.045 | -4.994 | 0.031 |
| 56928.12 | -5.506 | 0.018 | -6.952 | 0.017 | -6.711 | 0.018 | -5.694 | 0.048 | -4.830 | 0.033 |
| 56942.08 | -5.476 | 0.019 | -6.934 | 0.019 | -6.700 | 0.018 | -5.660 | 0.049 | -4.860 | 0.033 |
| 57282.14 | -5.451 | 0.019 | -6.931 | 0.016 | -6.694 | 0.014 | -5.778 | 0.019 | -5.031 | 0.034 |
| 57326.08 | -5.492 | 0.018 | -6.928 | 0.008 | -6.688 | 0.008 | -5.659 | 0.018 | -4.863 | 0.030 |
| 57347.00 | -5.479 | 0.013 | -6.939 | 0.009 | -6.693 | 0.009 | -5.695 | 0.018 | -4.934 | 0.029 |
| 57349.00 | -5.464 | 0.014 | -6.926 | 0.011 | -6.701 | 0.015 | -5.656 | 0.018 | -4.863 | 0.031 |

The maximum phase angle for Neptune is only 2 degrees so its effect on the planet's magnitude is negligible. However, the brightness of Neptune has increased over time as indicated by Equation A-8.1 (Schmude et al. 2016),

$$M_1(t) = C_0 + C_1 t$$

<div style="text-align: right">Equation A-8.1</div>

where *t* is the time in years measured from the reference year, 1984, which is the mid-time of the span of all Johnson-Cousins observations recorded for Neptune. The reason for this secular change is still not understood. The observed coefficients on the Johnson-Cousins system are listed in Table A-8.2 while the interpolated coefficients for the Sloan system are in Table A-8.3.

Table A-8.2. Observed Johnson-Cousins coefficients for Neptune

| C# | U | B | V | R | I |
|---|---|---|---|---|---|
| 0 | −6.380 | −6.550 | −6.940 | −6.510 | −5.360 |
| 1 | −0.00249 | −0.00249 | −0.00231 | −0.00610 | −0.00673 |

Table A-8.3 Interpolated Sloan coefficients for Neptune

| C# | u' | g' | r' | i' | z' |
|---|---|---|---|---|---|
| 0 | 0.0000 | 0.0000 | 0.0000 | 0.0000 | 0.0000 |
| 1 | −0.00249 | −0.00244 | −0.00399 | −0.00625 | −0.00671 |

As with Uranus, photometry in the near-IR may be problematic due to the methane bands which heavily blanket that portion of the Neptune's spectrum. Some of the observations obtained in the study by Schmude et al. (2016) were made with an I filter that was poorly matched to the Johnson-Cousins I-band response curve. This mismatch introduced uncertainty into the transformation to the standard Johnson-Cousins system. Therefore, the photometric I magnitude was omitted from Table 3 and subsequent analysis. The transformation from Johnson-Cousins R and I to Sloan z' is, likewise, also subject to considerable uncertainty. So, Table 5 does not contain a transformed Sloan magnitude. The $I_C$ and z' filters were well matched to the standard response curves and, therefore, their transformation uncertainties are small.

Like the other outer planets, fluxes used in the computation of synthesized magnitudes for Neptune were derived from the spectrophotometry of Karkoschka (1998). The resulting magnitudes were corrected for the difference between the year 1995, when the spectrophotometry was acquired, and the reference year.